\documentclass[aps,prb,twocolumn,10pt]{revtex4-2}
\usepackage{libertine}
\usepackage[T1]{fontenc}
\usepackage[libertine,upint]{newtxmath} 
\usepackage[cal=stixtwofancy,frak=stixtwo]{mathalfa}
\usepackage{graphicx,microtype,mathtools,siunitx,bm,xcolor}
\definecolor{linkscolor}{cmyk}{0.6, 0.3, 0, 0.9}
\usepackage[colorlinks=true,allcolors=linkscolor!60]{hyperref}

\DeclareSIUnit\angstrom{\text{Å}}

\DeclarePairedDelimiterX\braket[2]{\langle}{\rangle}{#1 \delimsize\vert #2}
\DeclarePairedDelimiterX\matrixel[3]{\langle}{\rangle}{#1 \delimsize\vert #2 \delimsize\vert #3}

\NewDocumentCommand{\grad}{e{_^}}{%
    \mathop{}\!
    \nabla
    \IfValueT{#1}{_{\!#1}}
    \IfValueT{#2}{^{#2}}
}

\DeclareMathOperator*{\SumInt}{%
\mathchoice%
  {\ooalign{$\displaystyle\sum$\cr\hidewidth$\displaystyle\int$\hidewidth\cr}}
  {\ooalign{\raisebox{.14\height}{\scalebox{.7}{$\textstyle\sum$}}\cr\hidewidth$\textstyle\int$\hidewidth\cr}}
  {\ooalign{\raisebox{.2\height}{\scalebox{.6}{$\scriptstyle\sum$}}\cr$\scriptstyle\int$\cr}}
  {\ooalign{\raisebox{.2\height}{\scalebox{.6}{$\scriptstyle\sum$}}\cr$\scriptstyle\int$\cr}}
}

\makeatletter
\g@addto@macro\@floatboxreset\centering
\makeatother

\begin{document}

\title{Magnetoelectric effect in the helical state of a superconductor/ferromagnet bilayer}

\author{V. Plastovets$^{1}$, A. Buzdin$^{1,2}$} 
\affiliation{$^{1}$University of Bordeaux, LOMA UMR-CNRS 5798, F-33405 Talence Cedex, France
\\
$^2$World-Class Research Center “Digital Biodesign and Personalized Healthcare”, Sechenov First Moscow State Medical University, 19991 Moscow, Russia}
\date{\today}

\begin{abstract}
    We study the microscopic mechanism of nucleation of the helical superconducting state maintained by a spin-splitting field and weak Rashba spin-orbit coupling in a 2D superconductor/ferromagnet system and present an exact solution for the Gor'kov equations, which provides a full description of the thermodynamic properties of the system under consideration. This permits us to obtain the Ginzburg-Landau expansion and use it to analyze the possibility of controlling magnetization using the transport supercurrent in both dc and quasistatic ac regimes. We show that these properties are related to the manifestation of the diode effect in superconducting systems with spin-orbit interaction.
\end{abstract}

\maketitle  

\section{Introduction}

The properties of superconductors with spin-orbit coupling (SOC) attracts a lot of attention due to the fact that SOC lifts the spin degeneracy and then generates triplet superconducting correlations (see, for example, \cite{bulaevskii1976superconductivity, edel1989characteristics, gor2001superconducting, amundsen2024colloquium}). 
Moreover, in the case of Rashba spin-orbit coupling (RSOC) the presence of Zeeman magnetic field leads to the appearance of the so-called helical phase with modulated superconducting order parameter $\Delta({\bf r})=\Delta e^{i{\bf q} {\bf r}}$, previously studied for the case of non-centrosymmetric crystals \cite{bauer2012non,smidman2017sc,PhysRevLett.94.137002,PhysRevB.78.224520,yanase2008helical,PhysRevB.78.144503}. The direction of the vector ${\bf q}\propto [{\bf h}\times {\bf n}]$ is determined by the cross product of the Zeeman field ${\bf h}$ and Rashba vector ${\bf n}$. At the phenomenological level of the Ginzburg-Landau (GL) theory, the helical state is associated with linear (known as Lifshitz invariant) and cubic in order parameter gradient terms \cite{edelstein1996ginzburg,PhysRevLett.75.2004}. In contrast with Fulde-Ferrell-Larkin-Ovchinnikov (FFLO) instability, which may emerge only at relatively low temperature and high magnetic field, the helical phase appears at arbitrary small Zeeman splitting. The properties of 2D superconductors with RSOC in parallel Zeeman field were extensively studied in Refs. \cite{PhysRevLett.89.227002,PhysRevB.75.064511,PhysRevB.76.014522,PhysRevB.92.014509,zwicknagl2017critical,PhysRevB.105.214508,PhysRevB.109.024516} mostly for the case of relatively strong SOC.

In this work we focus on the nucleation of the helical phase at arbitrary small values of RSOC for a model of a thin insulating ferromagnet film (FI) on top of a superconducting layer (S). We derive the corresponding GL functional up to the quadratic over the order parameter $\Delta$ terms straightforwardly from the microscopic Gorkov's equation. The helical (odd in ${\bf q}$) terms of the GL model coincide with that obtained diagrammatically in other works \cite{edel1989characteristics,PhysRevLett.75.2004,edelstein1996ginzburg,hasan2024supercurrent}. The appearance of the cubic gradient term of the order parameter gives rise to exciting phenomena such as intrinsic diode effect \cite{PhysRevLett.128.037001,kochan2023phenomenological,nadeem2023superconducting}, the characteristic features of which has been observed experimentally \cite{ando2020observation,pal2022josephson}. We demonstrate that the same cubic term provides the coupling of a superconducting current ${\bf j}$ in the S layer to the magnetization in the FI layer via effective magnetic field ${\bf H}_\text{eff}\propto [{\bf j}^3\times{\bf n}]$ and thus leads to a novel interesting magnetoelectric effect, which can, in some sense, be considered as an inverse diode effect.
As we will demonstrate, the process of magnetization switching via supercurrent operates on a fundamentally different principle compared to switching driven by spin-transfer torque in conventional spintronic devices.
This complements the known previous predictions about the generation of electron spin polarization by charge currents in semiconductors with strong SOC \cite{aronov1989observation,edelstein1990spin,ivchenko2008spin} and a similar effect in superconductors (Edelstein effect) \cite{PhysRevLett.75.2004,edelstein2005magnetoelectric}. 
The manipulation of the magnetic moment by a nondissipative supercurrent in both dc and ac regimes allows for the development of novel energetically efficient and fast operating (with times $\tau\sim T_c^{-1}$) device architectures for spintronics applications \cite{linder2015superconducting,mel2022superconducting}. 

The paper is organized as follows: we start by describing the microscopic model in Sec. \ref{SecII} and analyze its linearized GL form in Sec. \ref{SecIII}. Based on this, we then discuss the specifics of the current-magnetization coupling in the given system and its manifestation in Sec. \ref{SecIV}. Concluding remarks are given in Sec. \ref{SecV}.


\section{Model}\label{SecII}

We consider thin S/FI hybrid structure with thicknesses $d_S$ and $d_F$ respectively [see Fig. (\ref{fig1})]. The homogeneity of the order parameter over the S layer is ensured by $d_S\lesssim\xi(0)$, where $\xi(0)$ is the zero-temperature coherence length. The Zeeman spin-splitting field and the RSOC induced by the interface are assumed to be averaged over $d_S$ and thus be effectively uniform. We also focus on a weak RSOC regime, so there is only coupling of electrons with opposite spins and interband electron pairing is negligible. The suitable microscopic description for this effectively two-dimensional problem is given by the Gorkov's equations for normal $\hat{G}_{\omega_n}({\bf r},{\bf r}')$ and anomalous $\hat{F}_{\omega_n}({\bf r},{\bf r}')$ equilibrium Green's functions written in the spin space as
\begin{subequations} \label{Eq1}
    \begin{align}
\Big(i\omega_n - \hat{H}({\bf r}) \Big)\hat{G} \ + \ &\Delta({\bf r})i\hat{\sigma}_2\hat{F}^{\dagger} =\delta({\bf r}-{\bf r}'), \\ 
\Big(i\omega_n + \hat{H}^*({\bf r}) \Big)\hat{F}^\dagger \ - \ &\Delta^*({\bf r})i\hat{\sigma}_2\hat{G} =0.
    \end{align}
\end{subequations}
The external fields are included into the single-particle Hamiltonian
\begin{gather}
\hat{H}({\bf r}) = -\frac{{\nabla}^2}{2m} - \mu + {\bf h}{\bm \sigma} + \alpha({\bm \sigma}\times (-i)\nabla){\bf n},
\end{gather}
where ${\bf h}\propto {\bf M}$ is the Zeeman field induced by the magnetization in the FI layer; $\alpha$ and ${\bf n}$ are the strength and direction of RSOC; $\mu$ is the chemical potential; and ${\bm \sigma}$ is the vector of Pauli matrices. We use $k_B=\hbar=c=1$ hereinafter.   

The system (\ref{Eq1}) allows for an \textit{exact} solution for the case of a plane-wave modulation of the gap $\Delta({\bf r})=\Delta e^{i{\bf q}{\bf r}}$. Utilizing the momentum representation 
$\hat{G}({\bf r},{\bf r}')=\iint d{\bf p}d{\bf p}' e^{i{\bf p}{\bf r}-i{\bf p}'{\bf r}'}\hat{G}({\bf p},{\bf p}')$
we can write the answer as $\hat{G}\left({\bf p}, {\bf p}'\right) =\hat{G}\left({\bf p}\right)\delta\left({\bf p}- {\bf p}'\right)$ and $\hat{F}^{\dagger}\left({\bf p}, {\bf p}'\right) =\hat{F}^{\dagger}\left({\bf p}\right)\delta\left({\bf p}- {\bf p}'+{\bf q}\right)$,
where
\begin{subequations}\label{GF}
    \begin{align}
    \hat{G}\left({\bf p}\right)=\frac{ i\hat{\sigma}_2 \hat{\bar{G}}_0^{-1}\left({\bf p}-{\bf q}\right)}{ \Delta^2 i\hat{\sigma}_2 + \hat{{G}}_0^{-1}\left({\bf p}\right)i\hat{\sigma}_2\hat{\bar{G}}_0^{-1}\left({\bf p}-{\bf q}\right)  },
    \\ \label{ex_sol}
    \hat{F}^{\dagger}\left({\bf p}\right)=\frac{\Delta }{ \Delta^2 i\hat{\sigma}_2 + \hat{{G}}_0^{-1}\left({\bf p}\right)i\hat{\sigma}_2\hat{\bar{G}}_0^{-1}\left({\bf p}-{\bf q}\right)  }.
\end{align}
\end{subequations}
Here we defined the normal metal Green's function $\hat{G}_0({\bf p})=(i\omega_n - \hat{H}({\bf p}) )^{-1}$ and its conjugate $\hat{\bar{G}}_0({\bf p})=\hat{G}_0^*(-{\bf p})$; and the anomalous function $\hat{F}^{\dagger}\left({\bf p}, {\bf p}'\right)$ accounts for both singlet and triplet correlations in the superconductor with broken time-reversal and spatial inversion symmetries. 
Note that this is one of the rare cases where it is possible to find the exact solution in terms of Green functions for the superconducting system with non-trivial Hamiltonian. To our knowledge, these cases are limited to the superconducting state in helical \cite{bulaevskii1980helical} or conical magnets \cite{PhysRevB.99.024503}.
Analytically continued functions (\ref{GF}) contain all the information about electronic and thermodynamic properties of the modulated phase (see details in Supplemental Materials  \cite{SuppMat}\nocite{PhysRevLett.53.2437,abrikosov2012methods}).

The superconducting gap obeys the self-consistency equation
\begin{gather} \label{Eq2}
\Delta^*({\bf r}) = g  T\sum_{\omega_n} \text{Tr} \hat{\sigma}_+ \hat{F}^\dagger({\bf r},{\bf r}),
\end{gather}
where $g$ is the BCS coupling constant and $\hat{\sigma}_+=(\hat{\sigma}_x+i\hat{\sigma}_y)/2$. By substituting the solution (\ref{ex_sol}) we can cast Eq. (\ref{Eq2}) as $L^{-1}({\bf q})\Delta^*({\bf q})=0$, where the propagator reads as 
\begin{gather}\label{ker}
L^{-1}({\bf q})=\frac1g-T\SumInt_{\omega_n, {\bf p}} \text{Tr}  
\frac{\hat{\sigma}_+}{ \Delta^2 i\hat{\sigma}_2 + \hat{{G}}_0^{-1}\left({\bf p}\right)i\hat{\sigma}_2\hat{\bar{G}}_0^{-1}\left({\bf p}-{\bf q}\right)  }.
\end{gather}
%


\section{Linearized gap equation}\label{SecIII} 

To analyze the helical state near a critical line $T_c(h)$  \footnote{Well below the critical line the ground state can be determined by multiple modulation vectors forming the so-called stripe phase \cite{PhysRevB.76.014522,PhysRevB.109.024516}, which is not considered in the present work.}, separating normal and superconducting phases, it is sufficient to expand Eq. (\ref{ker}) up to linear in $\Delta$ terms, since we do not consider the changes of the equilibrium order parameter amplitude. Corresponding linearized operator $L^{-1}({\bf q})$ then simply reads in the symmetrized form as
\begin{gather}\label{L}
L^{-1}=\frac1g+T\SumInt_{\omega_n, {\bf p}} \text{Tr}  
\hat{\sigma}_+ \hat{\bar{G}}_{0}\left({\bf p}-\frac{{\bf q}}{2}\right)i\hat{\sigma}_2\hat{{G}}_0\left({\bf p}+\frac{{\bf q}}{2}\right).
\end{gather}
Choosing the particular configuration of the quantization axis as ${\bf h}||{\bf x}_0$ and ${\bf n}||{\bf z}_0$ we get  
\begin{gather}
\hat{G}_0({\bf p}) = \frac{1}{D_0}
\begin{pmatrix}
    i\omega_n-\xi_p & h-\alpha (p_y+ip_x) \\
    h-\alpha (p_y-ip_x) & i\omega_n-\xi_p
\end{pmatrix}, 
\end{gather}
where ${D_0=(i\omega_n-\xi_p)^2-h^2-\alpha^2 p^2+2\alpha h p_y}$; the normal metal spectrum is ${\xi_p}=p^2/2m-\mu$; and ${\bf p}=(p_x,p_y)$. The function $T_c(h)$ is given implicitly by the self-consistency equation as  $\text{max}_{\bf q}\big[L^{-1}({\bf q})=0\big]$, together with ${\bf q}(h)$ maximizing $T_c$.

\begin{figure}[] 
\begin{minipage}[h]{1.0\linewidth}
\includegraphics[width=1.0\textwidth]{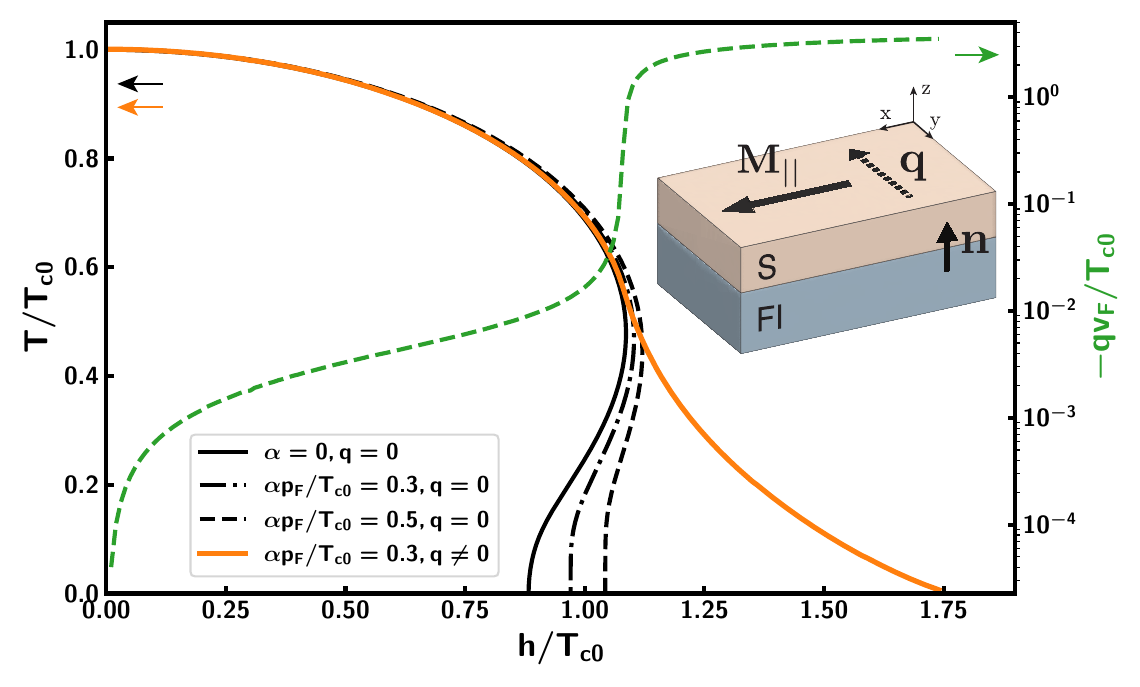} 
\end{minipage}
\caption{{\small Black lines: the critical temperature line $T_c(h)$ for the homogeneous state obtained from from Eq. (\ref{L}). Orange/Green line: $T_c(h)$ line and corresponding equilibrium modulation $q_y(h,T_c)$ interpolated between the perturbative solution Eq. (\ref{T_c}) for $h\ll T_{c0}$ and the numerical integration of Eq. (\ref{L}) for $\alpha=0$ (FFLO state). Insert: sketch of a S/FI system with vector ${\bf q}$ given by Eq. (\ref{q_eq}). 
}}
\label{fig1}
\end{figure}

First of all, we discuss the modifications of the phase diagram caused by both the bare RSOC generating triplet correlations, and the equilibrium modulation for (i) strong and (ii) weak Zeeman field:
(i) $L^{-1}(0)$ raises the $T_c(h)$ line and gives a noticeable shift of the zero-temperature paramagnetic limit ${h_{c0}=\pi T_{c0}e^{-\gamma}/2}$ [here $\gamma\approx0.577$] in the absence of the modulation, and for ${ \alpha p_F\ll h_{c0} \sim T_{c0} }$ we have 
\begin{gather}
    h_c\approx h_{c0} \left[ 1+\frac{5}{12}\frac{\alpha^2p_F^2}{h_{c0}^2}-\frac{\alpha^2 p_F^2}{2h_{c0}^2}\ln\left(\frac{\alpha p_F}{h_{c0}}\right) \right]. 
\end{gather}
Finite modulation ${\bf q}$ in turn additionally increases the critical field $h_c$, as can be seen from the numerical integration of Eq. (\ref{L}). We qualitatively demonstrate the dependence $T_c(h)$ for the helical state with $\alpha p_F\ll T_{c0}$ in Fig. \ref{fig1}. This can be obtained as an interpolation between the perturbative solution for $h\ll T_{c0}$ given by Eq. (\ref{T_c}) and the numerical integration of the function $\max_{{\bf q}}\big[ L^{-1}({\bf q})|_{\alpha=0}=0\big]$ for the FFLO state above the tricritical point  at $h>1.07 T_{c0}$.
As we mentioned, the helical state affects the $T_c(h)$ line for entire range of $h$, and it overlaps with the FFLO instability above the tricritical point with $q_\text{FFLO}\sim p_F T_{c0}/E_F$, making the standard first-order transition to the normal state a second-order one at low temperatures. The helical state here is almost indistinguishable from FFLO, since a non-zero RSOC formally leads to a small correction and overall $ q_\text{hel}\approx q_\text{FFLO}$.

(ii) For weak Zeeman field, namely $h\lesssim \alpha p_F\ll T_{c0}$, the helical state is more prominent. It possesses a small modulation vector $q_\text{hel}\ll \xi^{-1}(0)$ so that we can expand Eq. (\ref{L}) in its powers. Doing this and multiplying $L^{-1}$ by $|\Delta({\bf q})|^2$ we restore the GL free energy functional \cite{SuppMat} written in invariant form as
\begin{gather} \label{GL_q}
F_\text{S}=
\Big(a_0+ a_{h} {\bf h}^2 + {\bf q} [{\bf n}\times{\bf h}_{||}] \big(a_1+a_3 {\bf q}^2\big)+a_2{\bf q}^2  + a_4{\bf q}^4\Big) \Delta^2,
\end{gather}
where ${\bf h}_{||}$ is the in-plane component of the Zeeman field and the coefficients are 
\begin{gather} \notag
    a_0=\nu \ln\left(\frac{T}{T_{c0}} \right); ~ a_2 =\frac{7\zeta(3)v_F^2\nu}{32\pi^2T_{c0}^2}; ~  a_4 = \frac{-93\zeta(5)v_F^4\nu}{2048\pi^4 T_{c0}^4}; \\ \notag
    a_1 = \alpha C_\alpha \frac{ \zeta(5) \nu }{32\pi^4 T_{c0}^2}; \quad  a_3 =-   \alpha C_\alpha  \frac{1905\zeta(7) v_F^2\nu}{2048\pi^6 T_{c0}^4}; \\ \label{a_i}
    a_{h} = \nu \left(\frac{7\zeta(3)}{4\pi^2 T^2_{c0}} - \frac{31\zeta(5)}{32\pi^4}\frac{\alpha^2 p_F^2}{T^4_{c0}} \right)+\mathcal{O}\left(\frac{h^2}{T^2_{c0}}\right),
\end{gather}
where $C_\alpha=\alpha^2p_F^2/T_{c0}^2$.
The bare critical temperature is defined as ${T_{c0}=2\omega_D(\gamma/\pi) e^{-1/g\nu}+\mathcal{O}(\alpha^2p_F^2/E_F^2)}$ and it accounts for a small correction to the two-dimensional density of states $\nu=m/2\pi$ due to the RSOC. Equilibrium (preferred) modulation is fixed by a product of symmetry breaking fields:
\begin{gather}\label{q_eq} 
    {\bf q}_\text{eq} = -\frac{ a_1}{2 a_2}[{\bf n}\times {\bf h}_{||}],
\end{gather}
It provides the vanishing of a supercurrent in the ground state $\delta F_\text{S}/\delta {\bf q} =0$ and increases the critical temperature of the helical state. The total shift of $T_c$ up to $\mathcal{O}(h^4/T_{c0}^4)$ can be written as
\begin{gather}\label{T_c}
\nu\ln\left(\frac{T_c}{T_{c0}} \right) = -a_h {\bf h}^2 
+\frac{a_1^2 }{4 a_2}{\bf h}_{||}^2,
\end{gather}
where the first contribution comes from the isotropic suppression by Zeeman field, and the second one from the field component coupled to the vector ${\bf q}$. Eq. (\ref{T_c}) is in agreement with results from Ref. \cite{PhysRevLett.89.227002}. 
We shall make a small remark on the appearance of the coefficients $a_{1,3}$. The cubic dependence on $\alpha$ stems from the exact cancelling of the linear over $\alpha$ terms in 2D system for $\alpha p_F\ll T_{c0}$, which has also been found in previous works \cite{edel1989characteristics,PhysRevLett.75.2004,edelstein1996ginzburg,hasan2024supercurrent}. This may not be the case for the more general non-parabolic spectrum, where an order ${a_{1,3}\propto \alpha}$ can arise. 
Note that the linear in $\alpha$ terms also appears in the strong RSOC regime \cite{edelstein1996ginzburg,hasan2024supercurrent}. This limit is not described by our model since we neglect the splitting of the Fermi surface, but nevertheless we can qualitatively use the renormalization of $a_{1,3}$ by setting $C_\alpha \sim 1$.

The linear in gradient term associated with $\propto a_1 {\bf q}$ does not play any role in homogeneous systems, since it can always be gauged away from Eq. (\ref{GL_q}) (see for example \cite{PhysRevLett.128.037001}). The cubic term in turn is responsible for physically observable effects in superconductors, such as intrinsic diode effect \cite{wakatsuki2017nonreciprocal,he2022phenomenological,PhysRevLett.128.037001,kochan2023phenomenological,nadeem2023superconducting} and the new type of magnetoelectric effect.


\section{Magnetization control by supercurrent} \label{SecIV}

The coupling of the magnetic moment ${\bf M}$ of the ferromagnet to the modulation vector ${\bf q}$ opens up a way to magnetization control in the FI layer. Similar effects have been discussed theoretically in  systems of small (with lateral sizes $L_\text{FI}$ much less than the London penetration depth $\lambda_L$) ferromagnet nano-islands on a superconducting layer \cite{PhysRevB.95.100503}. This work describes the boundary physics related to the edge supercurrent generation \cite{PhysRevLett.115.116602,PhysRevLett.118.077001,olde2019superconducting}. The latter occurs thanks to the linear gradient term ($\propto {\bf q}$) in GL free energy, which can not be excluded by gauge transformation in inhomogeneous systems. We propose a very different mechanism working in the effectively macroscopic systems with $L_\text{FI}\gg \lambda_L$ and based on the diode effect, namely provided by the cubic term $\propto{\bf q}^3$ in Eq. (\ref{GL_q}).

The total free energy of the bilayer system integrated over its volume can be written as follows
\begin{gather}\label{GL_SF_total}
    \mathcal{F} = \Big(F_{SF} +F_{F}\Big)d_F,
\end{gather}
where $F_{F}$ is the free energy density of a specific configuration of the magnetic subsystem; and $F_{SF}$ is the energy of superconducting and magnetic interaction. Since the modulated phase with the fixed magnetization does not possess a charge current [Eq. (\ref{q_eq})], the application of an external one $\delta F_\text{S}/\delta {\bf q} = {\bf j}/2e$ may significantly alter the ground state through the direct ${\bf j}-{\bf M}$ interaction mediated by RSOC. This can be seen if one substitutes the corresponding modulation vector ${\bf q}\approx{\bf q}_\text{eq}+{\bf j}/4a_2e\Delta^2$ into Eq. (\ref{GL_q}). The interaction energy extracted from Eq. (\ref{GL_q}) can be written generally as $F_{SF}= -{\bf M}{\bf H}_\text{eff}$, where we effectively replace the supercurrent with a magnetic field defined in the lowest order in $\alpha$ as
\begin{gather}\label{H_eff}
    {\bf H}_\text{eff} = -\frac{\langle h\rangle}{M_0} \frac{   \ |a_3| [{\bf n}\times {\bf j}^3]}{ (4 a_2 e)^3 \Delta^4}\frac{d_S}{d_F}.
\end{gather}
Here we express the Zeeman field as ${\bf h}=\langle h\rangle{\bf M}/M_0$, where $\langle h\rangle$ is its averaged over $d_S$ value and $M_0$ is the magnetization magnitude.  Hereinafter we assume the modulus of $|{\bf M}|=M_0$ to be fixed, since the ferromagnet is in saturation regime justified by $T\sim T_{c0} \ll \Theta_K$. The effective field ${\bf H}_\text{eff}$ couples the applied supercurrent to the magnetic moment through the cubic term in Eq. (\ref{GL_q}). The linear coupling is naturally absent in Eq. (\ref{H_eff}) due to the gauge-invariance. However, according to Eq. (\ref{T_c}), the term $\propto a_1 {\bf q}$ in the expansion (\ref{GL_q}) provides an additional contribution to $F_{SF}$ responsible for the shift of $T_c$. The total energy of superconducting and magnetic interaction then reads 
\begin{gather}\label{F_S_M}
    F_{SF} = -{\bf M}{\bf H}_\text{eff} -   \frac{ a_1^2 \Delta^2 }{ 4 a_2} \frac{d_S}{d_F}\frac{\langle h\rangle^2}{M_0^2}   {\bf M}_{||}^2.
\end{gather}

This equation explicitly contains the magnetoelectric effect, and its connection to diode ${\bf q}^3$-effect is clear: the energy of a superconductor depends on the direction of the supercurrent for a fixed direction of magnetization orientation $F_{SF}({\bf j},{\bf M})\neq F_{SF}(-{\bf j},{\bf M})$ and vice-versa. This means that the free rotating magnetization vector can be adjusted by applied transport current, as the system will seek for the energy minimum. We empathize that phenomenological Eq. (\ref{F_S_M}) is valid for arbitrary values of $\alpha p_F$, and the second term in Eq. (\ref{F_S_M}) shows how the strong RSOC tends to create easy-plane magnetic anisotropy. Below we analyze the magnetization manipulation for two different configurations of the FI layer.

\subsection{Out-of-plane magnetization} \label{SecIVA}

\begin{figure}[] 
\begin{minipage}[h]{1.0\linewidth}
\includegraphics[width=1.0\textwidth]{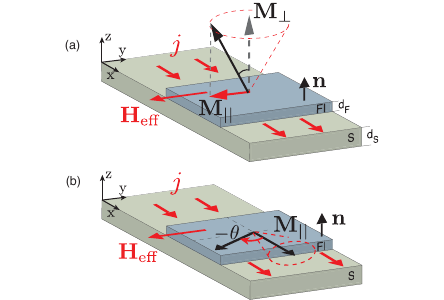} 
\end{minipage}
\caption{{\small Magnetoelectric effect controlled by the transport supercurrent ${\bf j}$ in the S/FI bilayer with $L_\text{FI}\gg \lambda_L$. In (a), the dc current ${\bf j}$ creates a tilt of the equilibrium out-of-plane magnetization along the effective field ${\bf H}_\text{eff}\propto [{\bf j}^3\times {\bf n}]$. In (b), the field ${\bf H}_\text{eff}$ rotates the magnetization away from the easy axis in the plane of FI. In both cases, the alternating current ${\bf j}(t)$ creates a precession shown by the dotted cones.
}}
\label{fig2}
\end{figure}

For the case of out-of-plane easy axis the moment is initially fixed as ${\bf M}_\bot=M_0 {\bf z}_0$ (see Fig. \ref{fig2}a), which corresponds to the zero equilibrium modulation [Eq. (\ref{q_eq})]. The applied dc transport current in the S layer induces the orthogonal in-plane component of the moment ${\bf M}_{||}$. To show this we introduce the magnetic anisotropy term 
\begin{gather}\label{GL_SF_A}
    \mathcal{F} = \Big[F_{SF} +A {\bf M}_{||}^2\Big]d_F
\end{gather}
with $A>0$. The effect of RSOC given by the second term in Eq. (\ref{F_S_M}) leads to a renormalization of the anisotropy parameter
\begin{gather}\label{A_eff}
    A_\text{eff} = A - \frac{ a_1^2 \Delta^2 }{ 4 a_2} \frac{d_S}{d_F} \frac{\langle h\rangle^2}{M_0^2} . 
\end{gather}
At $A_\text{eff}>0$ the equilibrium orientation of the magnetization remains out of plane and the supercurrent ${\bf j}$ creates a finite tilt of the moment along the field ${\bf H}_\text{eff}$, with the in-plane component
\begin{gather}\label{d_M}
    {\bf M}_{||} = {\bf H}_\text{eff}/2A_\text{eff} \propto -[{\bf n}\times {\bf j}^3].
\end{gather}
Created magnetization in turn induces the modulation with vector ${\bf q}_\text{eq}$ antiparallel to ${\bf j}$ according to Eq. (\ref{q_eq}). By increasing the applied current one can reach the complete $\pi/2$ rotation of the magnetic moment such that $|{\bf M}_{||}|=M_0$. The case of $A_\text{eff}<0$ will be considered below. 
Here we operate in terms of terminal (equilibrium) states of the magnetization. The transient rotation and relaxation of the magnetic moment is governed by Landau-Lifshitz-Gilbert (LLG) model, which gives us some characteristic switching time $\tau_{M}$. Thus, the manipulation of the moment can be realized either in the pulsed regime with $t\sim\tau_M$, or adiabatically at $t\gg\tau_M$. 

For an ac current ${\bf j}(t)$ that varies slowly on the timescale of superconducting relaxation time $\tau_\Delta\propto\hbar/(T_c-T)$, the superconducting system is quasistatic and ${\bf H}_\text{eff}$ is provided by equation (\ref{H_eff}) with ${\bf j}(t)$. Thus, we can show dynamic coupling of the low-frequency supercurrent ($\omega\ll\tau_\Delta$) and a homogeneous magnon mode ${\bf M}(t)$ within the framework of the model (\ref{GL_SF_A}). 
For illustrative purposes we consider the phenomenological LLG equation  
\begin{gather}\label{LLG}
    \frac{\partial {\bf M}}{\partial t} = -\gamma_1 \left[{\bf M}\times \left({\bf H}(t) - \frac{1}{d_F}\frac{\partial \mathcal{F}}{\partial {\bf M}_{||}} \right) \right]
    +\gamma_2 \left[{\bf M}\times \frac{\partial {\bf M}  }{\partial t}\right],
\end{gather}
where a small damping $\omega\gamma_2\ll\gamma_1$ is accounted.
The drive term in the R.H.S. is given by an external harmonic field ${\bf H}(t)$ and the effective field determined by the energy (\ref{GL_SF_A}). For small deviations of the in-plane moment $\delta {\bf M}_{||}\sim {\bf H}(t) \ll {\bf M}_\bot$ the linearized LLG equation reads 
\begin{gather}\label{LLG_lin}
    \left( \frac{\partial^2}{\partial t^2} + \omega_0^2 +2 \omega_0M_0\gamma_2  \frac{\partial}{\partial t} \right)
    \begin{pmatrix}
        \delta {\bf M}^{\bot j} \\
        \delta {\bf M}^{|| j}
    \end{pmatrix}
    \\ \notag
    =
    \frac{\omega_0}{2A_\text{eff}}
    \begin{pmatrix}
         \omega_0  &  \frac{\partial}{\partial t}   \\
         - \frac{\partial}{\partial t} & \omega_0
    \end{pmatrix}
    \begin{pmatrix}
         \left({\bf H}+{\bf H}_\text{eff}\right)^{\bot j} \\
         \left({\bf H}+{\bf H}_\text{eff}\right)^{|| j}
    \end{pmatrix},
\end{gather}
where the superscripts $\bot j, ||j$ mark components with respect to the applied current; the current induced field ${\bf H}_\text{eff}[{\bf j}(t)]$ is defined in Eq. (\ref{H_eff}); and we identified the resonant frequency $\omega_0  = 2 \gamma_1 M_0 A_\text{eff}$. Renormalization of the coefficient $A\rightarrow A_\text{eff}$ by the RSOC in the plane of the superconductor is responsible for the shift of ferromagnetic resonance ${\delta{\bf M}(\omega) \propto [\omega_0^2(T)-\omega^2+2\omega_0M_0\gamma_2i\omega]^{-1}}$ via the temperature dependent gap $\Delta(T)$. The field ${\bf H}_\text{eff}$ in turn produces an additional component to the drive $\propto j(t)^3$, which  can be generated by a transport current in the superconducting layer or by an electromagnetic radiation with typical for FMR microwave range. Strictly speaking, the drive of the magnetization by supercurrent requires the presence of dissipations in the superconducting system, which, in general, inevitably exist at finite frequencies at $T\approx T_c$ \cite{tinkham2004introduction}. Dissipative current can counteract the Gilbert damping through the spin-orbit coupling, but consideration of the corresponding nonequilibrium problem goes beyond the quasistatic approximation used in Eq. (\ref{LLG_lin}), and can be studied elsewhere.

We may also speculate that an inverse effect is possible, namely the excitation of a supercurrent by the induced magnetic moment precession. Such an effect is unattainable in the quasistatic (on the scale of the relaxation time $\tau_\Delta$) regime with $\partial_t \Delta \ll \Delta\tau_\Delta^{-1}$, where the magnetization ${\bf M}_{||}(t)$ will induce a quasi-equilibrium modulation vector (\ref{q_eq}), which always ensures the vanishing of a supercurrent. However, this constraint is naturally removed in an out-of-equilibrium state, where generally $\delta F_\text{S}/\delta {\bf q}\neq 0$.


\subsection{In-plane magnetization} \label{SecIVB}

For the case of in-plane easy axis the moment is initially fixed as $|{\bf M}_{||}|= M_0$  [Fig. \ref{fig2}b], which corresponds to nonzero equilibrium modulation (\ref{q_eq}). In the absence of the current the direction of the magnetization is either (i) isotropic or (ii) fixed by a crystal anisotropy. The applied dc transport current ${\bf j}$ can change the orientation of the moment, directing it along the field ${\bf H}_\text{eff}$. This will lead to (i) adjustable rotation of the magnetization or (ii) switching of the magnetization between different anisotropy axes. Let us introduce the total free energy as 
\begin{gather}\label{GL_SF_A_2}
    \mathcal{F} = \Big[ F_{SF} - B\left({\bf M}^4_x+ {\bf M}^4_y\right)\Big]d_F,
\end{gather}
where $B>0$; $F_{SF}$ is defined in Eq. (\ref{F_S_M}); and we assume a specific configuration of fourfold crystal symmetry determined by higher order terms. In order to consider the effect of anisotropy at $B\neq0$ we set the initial moment direction as ${\bf M}_{||}=M_0{\bf x}_0$. If the applied current ${\bf j}=j{\bf x}_0$ increases quasistatically we can neglect transient dynamics of the moment and find its equilibrium deviation from the $x$-axis parametrized by the angle $\theta$ (see Fig. \ref{fig2}b) with the energy
\begin{gather} \notag
    \mathcal{F} = \Big[ H_\text{eff} M_0 \sin\theta - B M_0^4\left(\sin^4\theta+\cos^4\theta\right)\Big] d_F.
\end{gather}
At $j\neq 0$ the initial state $\theta=0$ becomes metastable with respect to the stable $\theta=-\frac\pi 2$, and the magnetization rotation obeys the equation 
\begin{gather}
2\sin^3\theta-\sin\theta=\sqrt{2/27}H_\text{eff}/H^\text{cr}_\text{eff},
\end{gather}
where ${H_\text{eff}^\text{cr} = \frac{8}{3\sqrt{6}} \frac{B M_0^3 d_F}{d_S} }$ is the critical field corresponding to the instability point. At this field the applied current is strong enough to completely set the magnetization along the effective field ${\bf H}_\text{eff}$. The equilibrium rotation of the magnetization can be summarized as
\begin{gather}
    \theta = 
    \begin{cases}
        - \arcsin (\sqrt{2/27}H_\text{eff}/H^\text{cr}_\text{eff}) \quad &, H_\text{eff}\ll H_\text{eff}^\text{cr} \\
        -\pi/2 \quad &, H_\text{eff}\geqslant  H_\text{eff}^\text{cr}
    \end{cases}.
\end{gather}
For an isotropic case with $B=0$ the critical field is zero so that the magnetization always adjusts to the effective field ${\bf H}_\text{eff}$.

Magnon excitation in this configuration will be qualitatively similar to the previous case, albeit with some important remarks. Suppose the magnetization is fixed by anisotropy and the external drive causes its precession around the $x$-axis [Fig. (\ref{fig2}b)]. Deviation from an equilibrium can be accounted by adding terms $A_y{\bf M}_y^2+A_z{\bf M}_z^2$ to Eq. (\ref{GL_SF_A_2}). The parameter characterizing oscillations in plane is renormalized as $A_y\rightarrow A_y^\text{eff}$ according to Eq. (\ref{A_eff}), unlike the parameter $A_z$ for the out-of-plane dynamics. As a result this asymmetry leads to a different resonant frequency $\omega_0=2\gamma M_0 \sqrt{A_zA_y^\text{eff}}$. Note that the precession can again be excited by the supercurrent ${\bf j}(t)$ through the in-plane part of the magnetization.


\section{Discussion}\label{SecV} 

In order to ensure an uniform interplay of the two subsystems we suggest the following geometry for experimental observation of the magnetoelectric effect: A soft ferromagnetic film of permalloy or GdN, EuO, EuS ferromagnetic insulators with a thickness of the order of the magnetic correlation length $d_F \sim 10$ nm and a lateral size much exceeding $\lambda_L(0)$ is placed on a superconducting film of Nb, YBCO or FeSe with $d_S\sim \xi(0)\sim 100$ nm. The order of a magnitude of the effect can be quantitatively estimated via the induced field $H_\text{eff}$,  which can be enhanced in the limit of strong RSOC ($\alpha p_F\approx T_{c0}$) provided by inserting an additional Pt interlayer \cite{flokstra2023spin} and by a gate voltage, if additional top electrode is attached \cite{PhysRevLett.78.1335}. 
By choosing realistic parameters, such as $\alpha\sim 0.1v_F$; $q(j)\sim \xi^{-1}(0)$; $\langle h \rangle\sim 0.1 T_{c0}$; we can obtain the value of $H_\text{eff}\sim 10-100$ Oe, which is sufficient to provide the observable effects.
Since the magnetoelectric interaction is mediated by superconducting correlations, the field $H_\text{eff}$ is proportional to the condensation energy $\propto \Delta^2/E_F$, and therefore will be mostly pronounced in materials with relatively large $T_c/E_F$ ratio, such as Fe-based compounds \cite{kasahara2014field}. 

In conclusion, we discussed the nucleation of the helical phase in the bilayer S/F system for small SOC. Starting from the exact solution of the Gorkov's equations we derived the phenomenological theory of the magnetic moment control by transport supercurrent. The corresponding magnetoelectric effect originates from the cubic gradient term of the order parameter and has a common nature with the intrinsic diode effect in superconductors.

\acknowledgments
We thank A. Mel'nikov for fruitful discussions and M. Houzet for valuable comments. This work has been supported by GPR LIGHT and ANR SUPERFAST. A.B. acknowledges support by the Ministry of Science and Higher Education of the Russian Federation within the framework of state support for the creation and development of World-Class Research Center “Digital biodesign and personalized healthcare," No. 075-15-2022-304.

\newpage


\end{document}


\title{Supplemental Material}
\maketitle

\section{Exact Green function for superconductor with $\Delta({\bf q})$}

The normal and anomalous Green functions from Eq. (3) in the main text read in the momentum space as
\begin{gather}
    \hat{G}\left({\bf p}\right)=\frac{ i\hat{\sigma}_2 \hat{\bar{G}}_0^{-1}\left({\bf p}-{\bf q}\right)}{ \Delta^2 i\hat{\sigma}_2 + \hat{{G}}_0^{-1}\left({\bf p}\right)i\hat{\sigma}_2\hat{\bar{G}}_0^{-1}\left({\bf p}-{\bf q}\right)  }; 
    \quad \quad 
    \hat{F}^{\dagger}\left({\bf p}\right)=\frac{\Delta }{ \Delta^2 i\hat{\sigma}_2 + \hat{{G}}_0^{-1}\left({\bf p}\right)i\hat{\sigma}_2\hat{\bar{G}}_0^{-1}\left({\bf p}-{\bf q}\right)  }.
\end{gather}
Without loss of generality one can put $q_x=0$, since the helical state is associated with only $y$-axis in our setup. Below we write down the relevant matrix components of $\hat{G}\left({\bf p}\right)$ in a spin subspace:
\begin{gather}
    G_{\uparrow \uparrow}= G_{\downarrow \downarrow}= 
    \frac{-(i\omega_n+\xi_q)\Delta ^2 +(i\omega_n -\xi ) \left[(i\omega_n+\xi _q)^2 -h^2-2\alpha h \left(q_y-p_y\right)-\alpha^2  \left(p_x^2+\left(p_y-q_y\right){}^2\right)\right]}{D(i\omega_n)}, 
    \\
    F_{\uparrow \downarrow}^{\dagger}=-F_{\downarrow \uparrow}^{\dagger *}= 
    \Delta \frac{  -\Delta ^2+\left(h+\alpha(i p_x+  p_y)\right) \left(h+\alpha  \left(i p_x-p_y+q_y\right)\right)+(i\omega_n -\xi ) \left(i\omega_n+\xi _q \right)}{D(i\omega_n)}, 
\end{gather}
where $\xi_q=\xi({\bf p}-{\bf q})$ and denominator is
\begin{gather} \label{D}
    D(i\omega_n)=
    \left|\Delta ^2-(i\omega_n-\xi ) \left(i\omega_n+\xi _q \right)-\left(h+\alpha  p_y-i \alpha  p_x\right) \left(h-\alpha  \left(p_y-q_y\right)-i \alpha  p_x\right)\right|^2
    \\ \notag
    -\left|( i\omega_n-\xi ) \left(h-\alpha  \left(p_y-q_y\right)+i \alpha  p_x\right)+\left(i\omega_n+\xi _q \right) \left(h+\alpha  p_y-i \alpha  p_x\right)\right|^2.   
\end{gather}
Implementing analytical continuation as $i\omega_n \rightarrow \epsilon\pm i\Gamma$, where $\Gamma$ is the Dynes parameter \cite{PhysRevLett.53.2437}, we define retarted (R) and advanced (A) Green functions. From now on we assume the normal QP spectrum to be parabolic $\xi={\bf p}^2/2m-\mu$.

\section{Quasiparticle spectrum}

Spectrum of quasiparticle (QP) excitations corresponds to poles of $\hat{G}\left(\epsilon\right)$. In the above expressions the phase modulation vector ${\bf q}$ plays a role of a parameter, the equilibrium value of which should be found from the general principle of energy minimization. Suppose that we know the value  of ${\bf q}$ on the entire $T-h$ phase diagram. The spectral equation $D_G=0$ gives four branches and below we consider two limiting cases:

\

(i) The first one is well-known FFLO state which dominates for $h>h_\text{cr}$ and $T\ll T_{c0}$ with $|{\bf q}_\text{FFLO}|\sim p_F E_F/T_{c0}$. The RSOC gives a small correction to the modulation $q_\text{FFLO}+q_\text{hel}$ which is almost negligible. Thus, at $\alpha=0$ we obtain 
\begin{gather}
\epsilon_\text{FFLO}^0 =\pm h +\frac{{\bf v}{\bf q}}{2} - \frac{{\bf q}^2}{4m}\pm \sqrt{\Delta^2+ \left(\xi-\frac{{\bf v}{\bf q}}{2}+\frac{{\bf q}^2}{4m}\right)^2},
\end{gather}
where ${\bf v}=p/m(\cos\theta_p,\sin\theta_p)$.
This result formally resembles the one from Ref. \cite{bulaevskii1980helical}. Note that the spectral gap $\Delta_\text{spec}(\theta_p)=\text{min}_{|{\bf p}|}\epsilon({\bf p})$ in this case does not coincide with $\Delta$ and depends on the angle of the QP momentum with respect to the modulation vector. For strong exchange field the spectral gap vanishes for some $\theta_p$, which leads to the finite density of states (DOS) inside the energy region $|\epsilon-E_F|<\Delta$.

\

\begin{figure}[h] 
\begin{minipage}[h]{1.0\linewidth}
\includegraphics[width=0.48\textwidth]{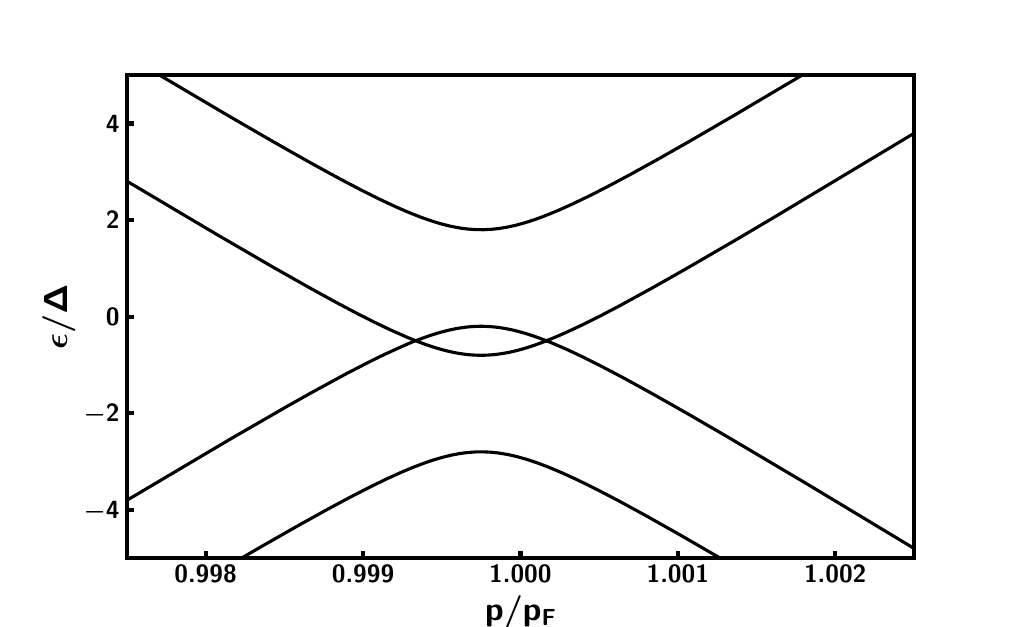} 
\includegraphics[width=0.48\textwidth]{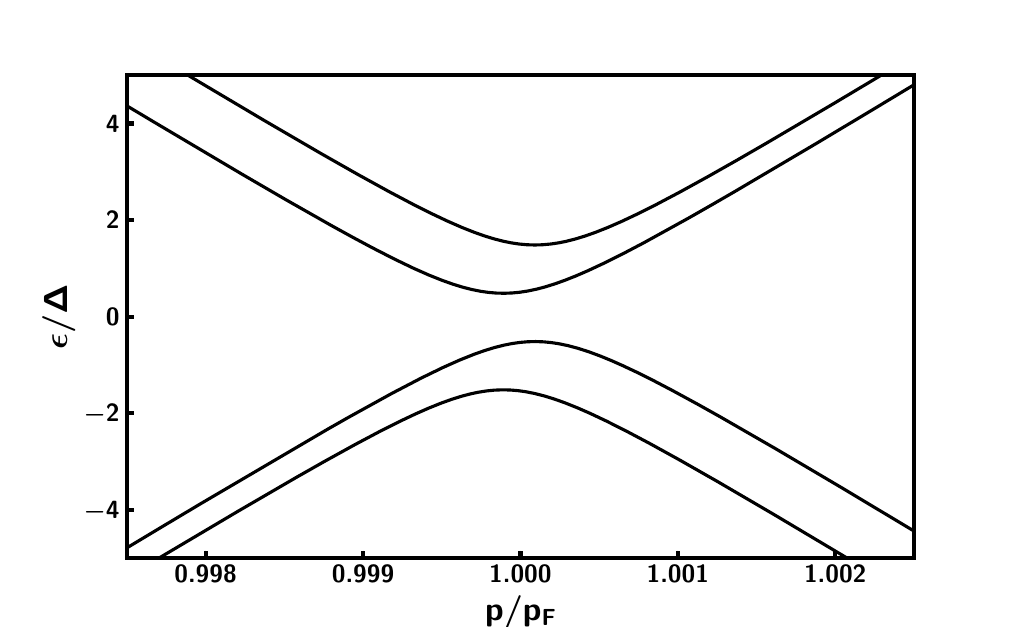} 
\end{minipage}
\caption{{\small Quasiparticle spectrum $\epsilon(|{\bf p}|, \theta_p=\pi)$ for: (a) FFLO state with $\alpha=0$, $h=1.3T_{c0}$, $q_y=0.5p_F T_{c0}/E_F$; (b) Helical state with $\alpha=0.2T_{c0}/p_F$, $h=0.5T_{c0}$, $q_y=0.8(\alpha p_F/T_{c0})^2 h/E_F$. For both plots $E_F/T_{c0}=1000$.
}}
\label{fig1}
\end{figure}

(ii) The second case is the helical state which is prominent at $h\ll T_{c0}$ and $T\approx T_{c0}$ with ${|{\bf q}_\text{hel}|\sim p_F (\alpha/v_F)(\alpha^2p_F^2/T_{c0}^2)(h/E_F)}$. The analytical expression for the spectrum is cumbersome even within the specified limits, so we provide illustrative plots in Fig. \ref{fig1}.

\vspace{1.0cm}

\section{Density Of States}

\begin{figure}[h] 
\begin{minipage}[h]{1.0\linewidth}
\includegraphics[width=0.48\textwidth]{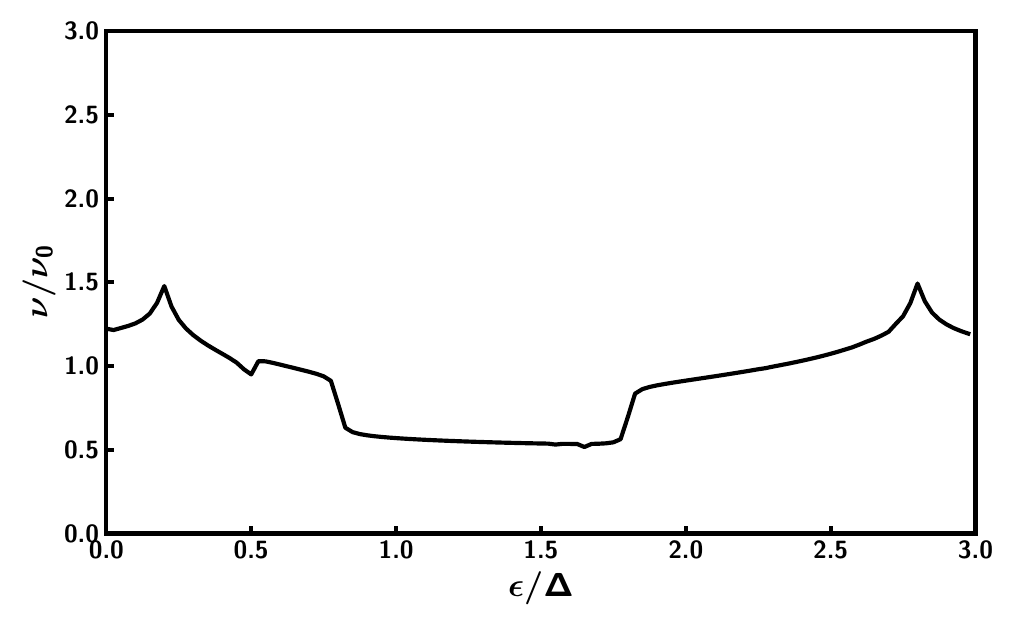} 
\includegraphics[width=0.48\textwidth]{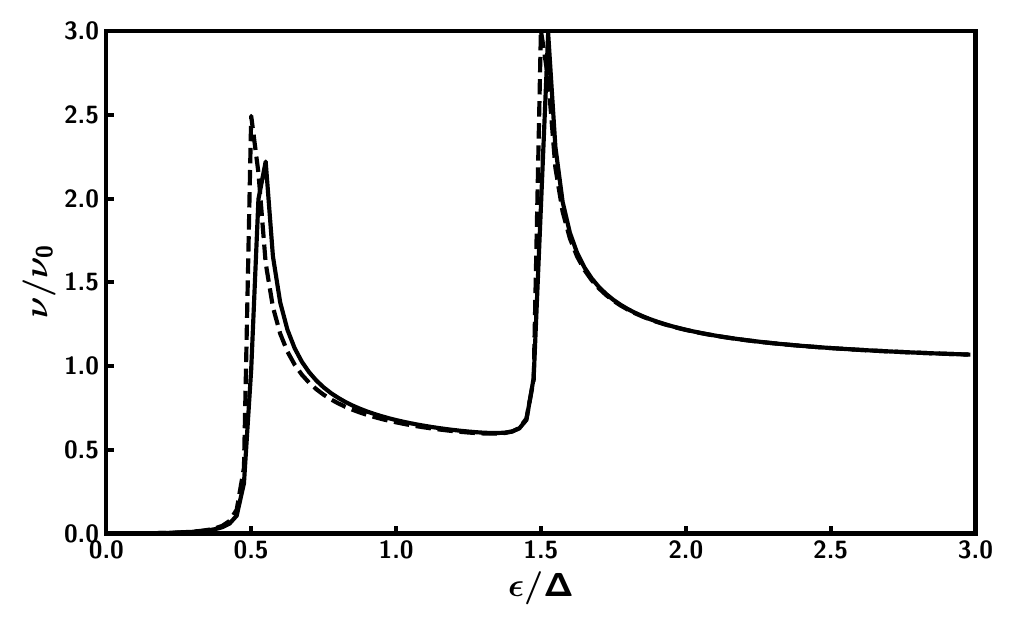} 
\end{minipage}
\caption{{\small Quasiparticle DOS for: (a) FFLO state with $\alpha=0$, $h=1.3T_{c0}$, $q_y=0.5p_FT_{c0}/E_F$; (b) Helical state with $\alpha=0.2T_{c0}/p_F$, $h=0.5T_{c0}$, $q_y=0.8(\alpha p_F/T_{c0})^2 h/E_F$ (DOS for $h=0.5T_{c0}$ and $\alpha=q_y=0$ is shown by a dashed line for a reference). For both plots $\Gamma=0.01T_{c0}$ and $E_F/T_{c0}=1000$.
}}
\label{fig2}
\end{figure}

The DOS can be calculated directly from the QP spectrum $\epsilon({\bf p})$: 
\begin{gather} 
    \frac{\nu(\epsilon)}{\nu_0}= \int \frac{d\theta_p}{2\pi}\frac{p}{m}\left(\frac{d\epsilon}{dp}\right)^{-1},
\end{gather}
where $\nu_0=m/2\pi \hbar^2$ is the 2D DOS in the normal state. On the other hand, the same result can be extracted directly from the retarded Green function (2):
\begin{gather} 
    \frac{\nu(\epsilon)}{\nu_0}
    = \int \frac{d\theta_p}{2\pi} \int dp \ \frac{p}{m\pi}\text{Im}G_{\uparrow\uparrow}^R({\bf p}).
\end{gather}
We will not discuss the angle-resolved DOS $\nu(\epsilon,\theta_p)$ focusing on the averaged result shown for FFLO and helical states in Fig. \ref{fig2}. One can notice the slight shift of two distinguished spin-split peaks caused by the finite vector ${\bf q}$ and associated energy ${\bf v}_F {\bf q}$ for the helical state. While for the FFLO state the picture is drastically different due to the gapless nature of the QP spectrum, and two peaks corresponding to the extrema of the spectral branches are clearly observed. These results complement the analysis made in Ref. \cite{PhysRevB.75.064511} within the quasiclassical approximation.

\section{Thermodynamic potential}

The QP spectrum (\ref{D}) allows one to obtain the ground state energy as
\begin{gather}
    \langle E \rangle = \int d\epsilon \ \epsilon \ \nu(\epsilon)  f(\epsilon) - \frac{\Delta^2}{g}, 
\end{gather}
where $f(\epsilon)$ is the QP distribution function. In the context of helical (or FFLO) state this could be potentially interesting to calculate the spin susceptibility $\chi_s(q) = \partial^2 \langle E \rangle/\partial h^2$ at low temperatures, however its numerical analysis is beyond the scope of this work.

~

Alternatively, it may be useful to have an expression for the thermodynamic potential $\Omega_s$ expressed in terms of the Green's function $\hat{F}^{\dagger}({\bf p})$. According to Ref. \cite{abrikosov2012methods} the quantity $\Omega_s$ can be expressed via the coupling constant $g$, which in turn is represented as a function of $\Delta$ in the self-consistency equation $L^{-1}({\bf q})\Delta^*({\bf q})=0$ (see Eq. (5) in the main text):
\begin{gather}
g^{-1}(\Delta, q)=T\SumInt_{\omega_n, {\bf p}} \text{Tr}  
\frac{\hat{\sigma}_+}{ \Delta^2 i\hat{\sigma}_2 + \hat{{G}}_0^{-1}\left({\bf p}\right)i\hat{\sigma}_2\hat{\bar{G}}_0^{-1}\left({\bf p}-{\bf q}\right)  },
\end{gather}
where $\SumInt_{\omega_n, {\bf p}}=\sum_{\omega_n}\int\frac{d^2p}{(2\pi)^2}$.
The expression for the thermodynamic potential  in the presence of spatial modulation should be modified and can be written as follows
\begin{gather}
\Omega_s-\Omega_n =  \int_0^\Delta \Delta'^2 \frac{d g^{-1}(\Delta',0)}{d \Delta'} d\Delta' + \Delta^2\Big[g^{-1}(\Delta,q)-g^{-1}(\Delta,0) \Big].
\end{gather}
Near the critical $T_c(h)$ line for $h,\alpha p_F\ll T_{c0}$ one can expand $\Omega_s-\Omega_n$ in powers of $\Delta$ and $q$. Using the identity 
$$\frac{7\zeta(3)\Delta^2(T\rightarrow T_c)}{32\pi^2T_c^2}\approx\ln\left(\frac{T_{c0}}{T}\right)$$
following from the self-consistency equation, we
thus obtain the standard Ginzburg-Landau expression for the free energy $F_S$ (see Eq. (9) in the main text).
The self-consistency equation written as $L^{-1}({\bf q})\Delta^*({\bf q})=0$ is generally given by a minimization of the energy over $\Delta({\bf q})$. In linearized form close to $T_c$ this gives the relation
\begin{gather}
\frac{\delta F_S}{\delta|\Delta({\bf q})|^2}=L^{-1}({\bf q}).
\end{gather}